\documentclass[referee]{cjaa}           

\usepackage{graphicx}                   
\usepackage{epsfig}

\begin{document}

   \title{Are radio-loud active galactic nuclei really follow the same $M_{\rm BH}-\sigma_{\rm *}$ relation for normal galaxies?  }

   \volnopage{Vol.0 (200x) No.0, 000--000}      
   \setcounter{page}{1}          

   \author{Yi Liu
      \inst{1,2}\mailto{}
   \and Dong Rong Jiang
      \inst{1}}
   \offprints{Yi Liu}                   

   \institute{Shanghai Astronomical Observatory, CAS, Shanghai 200030, China\\
             \email{yliu@shao.ac.cn}
        \and
                  Graduate School of the Chinese Academy of
Sciences, BeiJing 100039, China}

   \date{Received~~ month day; accepted~~ ~~month day}

\abstract{In order to examine the relationship between the black
hole mass $M_{\rm BH}$ and stellar velocity dispersion $\sigma_{\rm
*}$ in radio-loud active galactic nuclei (AGNs), we study two
effects which may cause the uncertainty of black hole mass estimates
for radio-loud AGNs: the relativistic beaming effect on the observed
optical continuum radiation and the orientation effect on broad
emission line width. After correcting these two effects in black
hole mass calculations, we re-examine the $M_{\rm BH}-\sigma_{\rm
[OIII]}$ relation for a sample of radio-loud and radio-quiet AGNs
compiled from the literatures, and find the $M_{\rm BH}-\sigma_{\rm
[OIII]}$ relation in radio-loud AGNs still deviate from that in
nearby normal galaxies and radio-quiet AGNs. We also find there is
no significant correlation between radio jet power and narrow [OIII]
line width, indicating no strong interaction between radio jet and
narrow line region. The deviation from $M_{\rm BH}-\sigma_{\rm *}$
relation in radio-loud AGNs may be intrinsic, or the [OIII] line
width is not a good indicator of $\sigma_{\rm *}$ for radio-loud
AGNs.
   }

   \authorrunning{Yi Liu \and Dong Rong Jiang }            
   \titlerunning{Black hole mass and narrow [OIII] line width}  

\maketitle

%
%
\section{Introduction}           
\label{sect:intro}
There is abundant evidence show that the evolution of black holes
and that of their host galaxies appears to be closely coupled.
Kormendy \& Richstone (1995) and Magorrian et al. (1998) showed that
the central black hole mass correlates with the bulge mass and
luminosity. The stellar velocity dispersion $\sigma_{\rm *}$ in the
galactic bulge is also related with the mass of the center black
hole (the $M_{\rm BH}-\sigma_{\rm *}$ relation). Gebhardt et al
(2000a) and Ferrarese \& Merritt (2000) found that the correlation
between $M_{\rm BH}$ and $\sigma_{\rm *}$ is strong, suggesting a
link between the formation of the bulge and the black hole. After
investigating 31 nearby inactive galaxies, Tremaine et al. (2002)
presented a relation as:
\begin{equation}
M_{\rm BH}=10^{8.13}(\sigma_{*}/(200 kms^{-1}))^{4.02} M_{\rm
\odot}.
 \end{equation}
However, the bulge stellar velocity dispersion $\sigma_{\rm *}$ in
the QSO is generally difficult to measure directly. Nelson \&
Whittle (1995; 1996) made a comparison of bulge magnitudes, [OIII]
line widths, and $\sigma_{\rm *}$ in Seyfert galaxies, and found a
good agreement between $\sigma_{\rm *}$ and $\sigma_{\rm [OIII]}$ ($
\sigma_{\rm [OIII]}=FWHM([OIII])/2.35$) in statistical sense. The
above relationship for normal galaxies also holds for active
galaxies (Gebhardt et al. 2000b; Ferrarese et al. 2001; Wang \& Lu
2001; Boroson 2003; Shields et al. 2003). The $M_{\rm
BH}-\sigma_{\rm [OIII]}$ relation for all their radio-quiet sample
follow that derived from nearby normal galaxies. It provides us a
method to calculate the mass of black holes differ from that in
Kaspi et al. (2000), which derived from active galactic nucleus
broad emission line width and continuum luminosity.

However, Bian \& Zhao (2004) presented a distinct result that the
relationship between the mass of black hole and velocity dispersion
in narrow line Seyfert1s and radio-loud AGNs deviates from the
relationship which in radio-quiet AGNs and nearby normal galaxies.
They suggested that the deviation in radio-loud AGNs from the
$M_{\rm BH}-\sigma\rm_{*}$ relation might be due to the measurements
uncertainties of $\sigma_{\rm [OIII]}$ or $M_{\rm BH}$. Bonning et
al. (2005) tested the use of [OIII] line widths as a surrogate for
$\sigma_{\rm *}$ by studying the $M_{\rm HOST}-\sigma_{\rm [OIII]}$
relation in a sample of quasars for which the host galaxies
luminosity has been measured. They found an increase of $\sigma_{\rm
[OIII]}$ with $\sigma_{\rm *}$ in covering a wide range of measured
or inferred $\sigma_{\rm *}$, though the radio-loud AGNs have
$\sigma_{\rm [OIII]}$ smaller by 0.1 dex than radio-quiet QSOs of
similar $L_{\rm HOST}$. Greene \& Ho (2005) compare $\sigma_{\rm *}$
with the widths of several narrow emission lines in a sample of
narrow-line Seyfert galaxies from Sloan Digital Sky Survey (SDSS)
and found that $\sigma_{\rm [OIII]}$ exceeds $\sigma_{\rm *}$ about
0.13 dex. However, Salviander et al. (2006) found that $\sigma_{\rm
[OIII]}$ agrees with $\sigma_{\rm [OII]}$ well in their sample of
SDSS QSOs. Bian, Yuan \& Zhao (2005) investigated the radial
velocity difference between the narrow emission-line components of
[OIII] in a sample of 150 SDSS NLS1 galaxies, and found that profile
of [OIII] indicating two kinematically and physically distinct
regions. The [OIII] line width depends not only on the bulge stellar
gravitational potential, but also on the central black hole
potential. Moreover, the interaction between the radio jets and the
narrow line region (NLR), or radio jets inspire the star formation
might also influence the physical and dynamical state of the [OIII]
line width and its intensity.

Kaspi et al. (2000) derived an empirical relationship between the
broad line region (BLR) size and the optical continuum luminosity at
5100$\rm {\AA}$ using the reverberation mapping technique. The empirical
relationship has been frequently adopted to estimate the BLR size
and then to derive the black hole mass for AGNs samples. However,
the relativistic jets of radio-loud AGNs both dominate the radio
radiation and contribute significantly to the optical luminosity.
The black hole mass in radio-loud AGNs would be overestimated by
using the empirical relationship between the BLR size and optical
luminosity at 5100$\rm {\AA}$ which is obtained from the sample of
radio-quiet AGNs (Kaspi et al. 2000). On the other hand, as the jet
axes are near to the line of sight for radio-loud AGNs, the
geometrical effects might affect the observed widths of the broad
$H\beta$ emission line and hence the black hole masses in radio-loud
AGNs. These uncertainties in black hole mass calculation might
influence the $M_{\rm BH}-\sigma_{\rm [OIII]}$ relation in
radio-loud AGNs. In order to eliminating the beaming effect in
optical continuum radiation for radio-loud AGNs, Wu et al. (2004)
provided a method to calculate the black hole mass by a tight
empirical relationship between the BLR size and the $H\beta$
emission line luminosity. In this paper, we re-compare the $M_{\rm
BH}-\sigma_{\rm [OIII]}$ relation in a sample of radio-loud and
radio-quiet AGNs after correcting the uncertainty for black hole
mass, and then examine whether the $\sigma_{\rm [OIII]}$ traces the
$\sigma_{\rm *}$ in radio-loud AGNs. The relationship between the
radio jet power and the [OIII] line width also be investigated. We
used a cosmology with $H_{0}=70 {~km ~s^ {-1}~Mpc^{-1}}$,
$\Omega_{M}=0.3$, $\Omega_{\Lambda} = 0.7$. All values of luminosity
used in this paper are corrected to our adopted cosmological
parameters.


\section{Sample and Method}
\label{sect:data}
We started this work with the sample of Xu et al. (1999) which has
409 sources from the literature or through the NASA Extragalactic
Database (NED). To estimate the black hole mass and $\sigma_{\rm
[OIII]}$, we then searched the literature for all available
measurements of the full width of half maximum (FWHM) of both the
broad $H\beta$ and narrow line $[OIII]_{\rm 5007}$, as well as of
the line flux of $H\beta$. Finally, a sample of 123 AGNs was
constructed, of which 25 are flat-spectrum radio-loud AGNs (FS), 35
are steep-spectrum (SS) radio-loud AGNs and 63 radio-quiet AGNs. The
number of radio-loud AGNs in this sample is similar to that in Bian
\& Zhao (2004).

Table 1 lists our sample with the relevant information for each
object. Columns (1)-(3) give the object's IAU name, radio spectrum
type (FS denotes flat-spectrum and SS denotes steep-spectrum
radio-loud AGNs, and RQ shows the radio-quiet AGNs respectively) and
redshift. In columns (4)-(6), we list the black hole mass, the
references for adopted FWHM of broad $H\beta$ emission line, and the
references for adopted luminosity of broad line $H\beta$. In columns
(7)-(8), we present the $\sigma_{\rm [OIII]}$ and its references. A
more detailed notation is given at the caption of Table 1.

Through the reverberation mapping method, Kaspi et al.(2000)
presented an experiential relationship between the BLR size and the
monochromatic luminosity at 5100 $\rm {\AA}$, and then the mass of the
black holes can be estimated by using ${M_{\rm BH}=R_{\rm
BLR}V^2G^{-1}}$. However, in radio-loud AGNs, there are two effects
would cause the uncertainty of black hole mass estimate. Firstly,
the radio observations shown that the orientation of the jets in the
radio-loud AGNs, especially in the flat spectrum radio-loud AGNs,
are close to the line of sight. The synchrotron emission of the jet
are then Doppler boosted, and the contribution of the
synchronization emission at the optical band may dominate over the
thermal emission from the disk. So the mass of black hole in
radio-loud AGNs could be overestimated. The second effect may cause
the uncertainty of the black hole mass estimate is: if the geometry
of the BLR is disk-like, the broad line width will depend on the
orientation of the disk. The broad line widths of the radio-loud
AGNs will be smaller than those in radio-quiet AGNs, even if they
have similar black hole masses and BLR sizes (see Cao 2000; Xu \&
Cao 2006 for BL Lac objects). Wu et al. (2004) gave the relationship
between the broad emission line $H\beta$ luminosity and size of
broad line region which is derived from reverberation mapping
method. Then the optical luminosity eliminated beaming effect can be
calculated by using the broad line emission luminosity. As for the
orientational effect for emission line width, Lacy et al. (2001) put
forward using $R_{\rm c}^{0.1}$ ($R_{\rm c}$ is the ratio of the
core to the extended radio luminosity in the rest frame of the
sources) factor to modify this effect, i.e for all radio-loud AGNs.
However, this modification is invalid for the sources with $0<R_{\rm
c}<1$ and we only correct this orientation effect for the sources
with $R_{\rm c}>1$. After these two corrections, we then calculated
the mass of black hole in our sample and these compared the $M_{\rm
BH}-\sigma_{\rm [OIII]}$ relation in radio-loud and radio-quiet
AGNs. The results are plotted in Fig. 1.

\section{Results and Discussion}
\label{sect:analysis} In Fig. 1(a)-(c) we plot the relationship
between $\sigma_{\rm [OIII]}$ and black hole mass $M_{\rm BH}$ from
different methods, respectively.  The black hole masses in Fig. 1(a)
are estimated without any corrections. Fig. 1(b) shows the black
hole mass with beaming effect correction but without orientation
effect correction. In Fig. 1(c), the black hole masses are estimated
with both beaming effect (for all radio-loud objects) and
orientation effect (for only flat spectrum radio-loud AGNs)
corrections. The solid circles denote flat spectrum radio-loud AGNs;
the open circles present steep spectrum radio-loud AGNs, and the
stars are for radio-quiet AGNs. The solid lines plotted in Fig.
1(a)-(c) show the $M_{\rm BH}-\sigma_{\rm *}$ relation defined by
Tremaine et al. (2002).

\clearpage
\begin{figure}
   \vspace{2mm}
   \begin{center}
   \hspace{3mm}\epsfig{figure=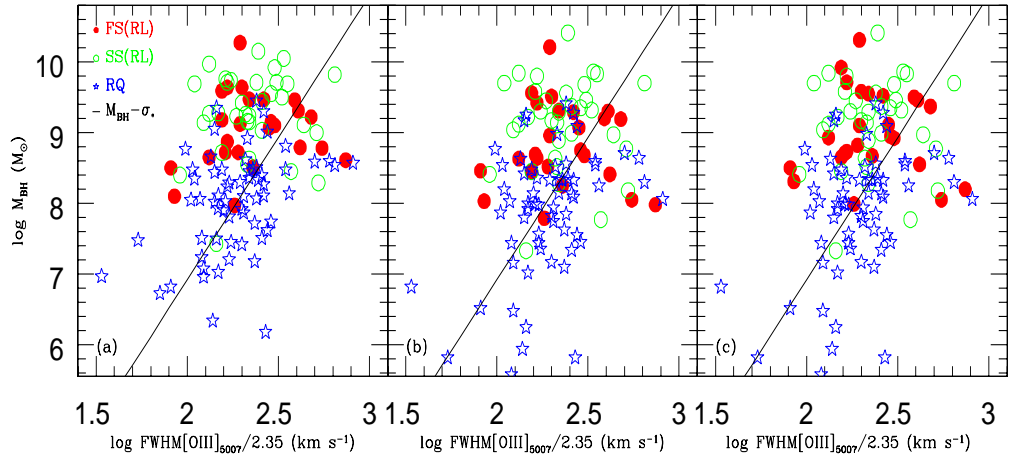,width=150mm,height=180mm,angle=0.0}
   \caption{Black hole mass versus $\sigma_{\rm OIII}$ for our sample.
   (a): Black hole mass derived from $H\beta$ line width without
   orientation correction and optical continuum luminosity. (b): Black
   hole mass derived from $H\beta$ line width without orientation
   correction and $H\beta$ luminosity. (c): Black hole mass estimated
   by correcting both the Doppler beaming effect and orientation effect.
   Solid circles denote the flat spectrum radio-loud AGNs;
open circles show the steep spectrum radio-loud AGNs and pentagons
plot the radio-quiet AGNs. The solid line shows $M_{\rm
BH}-\sigma_{\rm *}$ relation from equation (1). }
   \label{Fig:lightcurve-ADAri1}
   \end{center}
\end{figure}
\clearpage

The beaming and orientation effects cause the black hole mass to
move towards different opposite positions. On the one hand, similar
as radio luminosity, the optical luminosity of radio-loud quasars
can also be contaminated by the relativistically beamed jet
emission. In fact, the optical emission of radio-loud quasars is a
mixture of thermal and non-thermal emission, and tends to be
orientated with their jets beamed along our line of sight, although
not explicitly applicable on a source-by-source basis. Thus the
black hole mass in radio-loud AGNs would be reduced after correcting
the beaming effect for radio jet (Fig. 1(a) and Fig. 1(b)). On the
other hand, if the geometry of the broad line region is disk-like,
the object has smallest line width when the jet axis is along with
the line of sight. After correcting for orientation effect, the
black hole mass will be increased.

Figure 1(c) shows that the radio-quiet and radio-loud AGNs occupy
two distinct regions in the $M_{\rm BH}$-$\sigma_{\rm [OIII]}$
plane. As the same results of other authors, the radio-quiet AGNs
follow that relationship for our sample, but the radio-loud AGNs
still deviate the $M_{\rm BH}-\sigma_{\rm *}$ relation of nearby
normal galaxies, although the beaming effect and orientation effect
in black hole mass calculation are eliminated. The black hole mass
of radio-loud AGNs plotted in Fig. 1(c) are calculated by the width
of $H\beta$ line and the luminosity of $H\beta$ which has been
described in section 2. The beaming and orientation effect are two
factors which will cause uncertainties of black hole mass. In order
to eliminate the contamination by the emission from radio jets,
continuum luminosity at 5100$\rm {\AA}$ was substituted by the broad line
emission. However, the radio-loud AGNs remain the $M_{\rm
BH}-\sigma_{*}$ relation deviation after removing orientation
factor. But we must keep in mind that the relationship between the
broad line region radii $R_{\rm BLR}$ and broad emission line
luminosity $L_{\rm H\beta}$ which we used to calculate the black
hole mass in our sample was derived from a sample for radio-quiet
AGNs (Kaspi et al. 2000). Until now, the $R_{\rm BLR}~-~L_{\rm
5100\rm {\AA}}$ relation established from reverberation mapping method on
radio-loud AGNs is still unavailable. Thus we are not clear whether
the relationship between the $R_{\rm BLR}$ and $L_{\rm 5100\rm {\AA}}$ or
$L_{\rm H\beta}$ for radio-loud AGNs is systematically different
from that of radio-quiet AGNs. However, if this difference in
radio-loud AGNs does exist, that means the $R_{\rm BLR}$ in
radio-loud AGNs will one order of magnitude lower than that one in
radio-quiet AGNs for the similar optical/line luminosity since the
broad line width $\rm FWHM_{\rm H\beta}$ is similar for radio-loud
and radio-quiet AGNs. However, the evidence for this kind of
difference is not found so far, thus we think the uncertainties in
black hole mass calculation might not be the main reason which leads
this deviation in radio-loud AGNs.


Assuming radio-loud AGNs to follow the same $M_{\rm BH}-\sigma_{\rm
*}$ relation defined by nearby normal galaxies and radio-quiet AGNs,
we can derive the [OIII] line width which they should be from the
$M_{\rm BH}-\sigma_{\rm [OIII]}$ relation and then derive the
deviation to observed [OIII] line width($\rm \Delta[OIII]$). Figure
2 and Fig. 3 present the distribution of [OIII] line width and the
distribution of $\rm \Delta[OIII]$ for radio-loud and radio-quiet
AGNs, respectively. The medians of [OIII] line width for these two
sub-samples are similar (510 $\rm km~s^{-1}$ for radio-loud AGNs and
480 $\rm km~s^{-1}$ for radio-quiet AGNs). However, it is obvious
that the observed [OIII] line widths in radio-loud AGNs are smaller
than those expected by the $M_{\rm BH}-\sigma_{\rm *}$ relation for
normal galaxies.


\clearpage
\begin{figure}
   \vspace{2mm}
   \begin{center}
   \hspace{3mm}\epsfig{figure=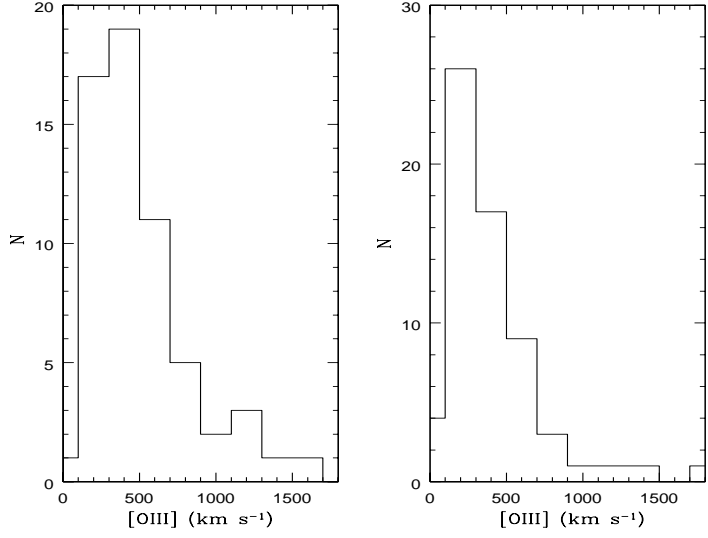,width=100mm,height=80mm,angle=0.0}
   \caption{Distribution of observed [OIII] line width for radio-loud AGNs (left) and radio-quiet AGNs (right). }
   \label{Fig:lightcurve-ADAri2}
   \end{center}
\end{figure}

\begin{figure}
   \vspace{2mm}
   \begin{center}
   \hspace{3mm}\epsfig{figure=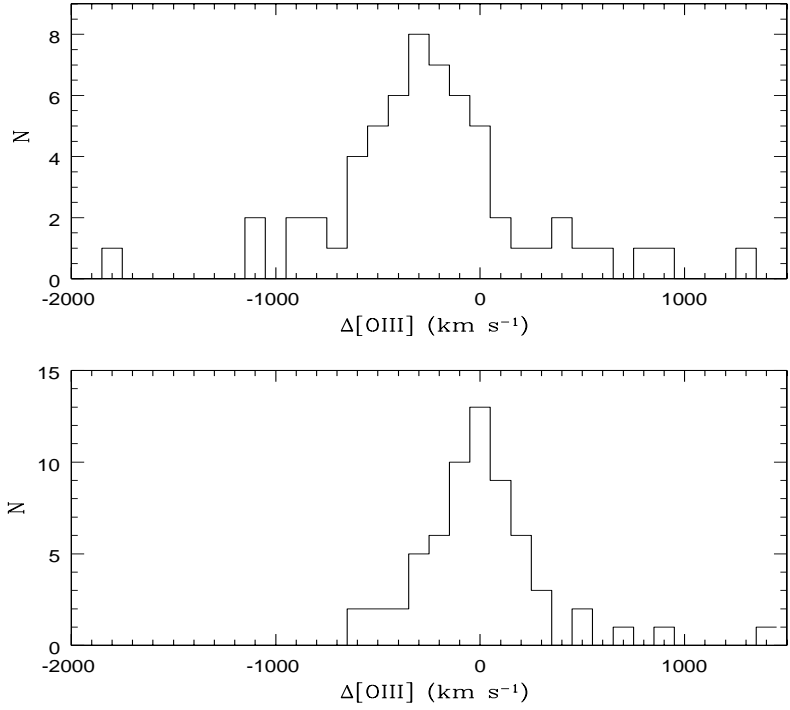,width=120mm,height=100mm,angle=0.0}
   \caption{Distribution of difference between the observed [OIII] line widths and expected by
   $M_{\rm BH}-\sigma_{\rm *}$ relation for radio-loud AGNs (upper) and radio-quiet AGNs (lower). }
   \label{Fig:lightcurve-ADAri3}
   \end{center}
\end{figure}
\clearpage

Although several groups investigated whether the $\sigma_{\rm
[OIII]}$ traces the $\sigma_{\rm *}$, the use of $\sigma_{\rm
[OIII]}$ for $\sigma_{\rm *}$ is controversial. The outflow combined
with extinction of the far side of the NLR might be as the result
which caused asymmetry and non-Gaussian profile for [OIII] emission
line (Nelson \& Whittle 1995). Other strong iron emission lines,
like FeII line, locates close to the [OIII] line may obscure them.
There are some evidences suggesting that radio jets influence the
physical and dynamical state of the NLR (de Bruyn \& Wilson 1978; Ho
\& peng 2001). Nelson \& Whittle (1996) found that the strong radio
sources have broader [OIII] line width than weak radio sources. It
seems that the jet plasma interacts with and accelerates the NLR,
thus boosting the line width. However, radio-loud QSOs on average
have smaller [OIII] line width than radio-quiet QSOs was found by
Bonning et al. (2005). In the study for Smith et al. (1990), they
found that no superviral [OIII] line widths among that objects with
powerful radio jets at kpc scale.

The radio jet power, as a fundamental radio parameter indicating the
energy transported through the radio jet from the central engine,
can be used to investigate the relationship between the radio jet
and the narrow line regions. We used the formula derived from Punsly
(2005):
\begin{equation}
Q_{\rm jet}=5.7\times10^{44}(1+z)^{1+\alpha}Z^{2}F_{\rm 151}  ~~
erg~s^{-1}
\end{equation}
\begin{equation}
Z\approx3.31-3.65\times{[(1+z)^{4}-0.203(1+z)^{3}+0.749(1+z)^{2}+0.444(1+z)+0.205]^{-0.125}}
\end{equation}
to estimate the jet power, where $F_{\rm 151}$ is the optically thin
flux density from the lobes measured at 151 MHz in units of Janskys,
and the value of $\alpha \approx 1$ is suggested by the observations
(Kellermann, PaulinyToth \& Williams 1969) as a good fiducial value
(see Punsly 2005 for more details). We also estimated the jet power
by substituting the extrapolated extended 151 MHz flux density
($\alpha=1.0$), instead of the measured 151 MHz flux since the 151
MHz emission can be from radio cores with Doppler boosted effects
(Liu, Jiang \& Gu. 2006). The relationship between the radio jet
power and $\sigma_{\rm [OIII]}$ is shown in Fig. 4, the solid
circles denote all radio-loud AGNs in our sample, the open squares
added on solid circles present objects with black hole mass in range
from $10^{8.5}M_{\odot}$ to $10^{9.5}M_{\odot}$. Using the Spearman
rank correlation analysis, we find a significant correlation between
radio jet power and [OIII] line width with a correlation coefficient
of $r=0.32$ at $\gg99$ per cent confidence. It should be noted with
caution that this correlation may be caused by the common dependence
of black hole mass. It may not be an intrinsic correlation, thus we
check the correlation between the $\rm \Delta[OIII]$ line width and
radio jet power. Figure 5 shows the relationship between $\rm
\Delta[OIII]$ and radio jet power in our sample. However, no
significant correlation is present between $\rm \Delta[OIII]$ line
width and radio jet power. It may support the scenario that there is
no strong interaction between radio jet and narrow line region.
Moreover, it should be cautious if there exists the interaction
between the radio jet and narrow line region, this effect may
broaden the narrow line width, and the $^{'}\rm true^{'}$ [OIII]
width will be narrower than that observed, which leads to enlarge
the offset in $M_{\rm BH}-\sigma_{\rm [OIII]}$ for radio-loud AGNs.
Greene \& Ho (2005) found the presence of core radio emission seems
to have no impact on the observed [OIII] line width, and we derive
same result by using radio jet power. However, they found the
extended radio sources appear to have narrower line width, as well
as in our sample. In spite of the offset in the $M_{\rm
BH}-\sigma_{\rm [OIII]}$ relation is unknown, Bonning et al. (2005)
pointed out the radio-loud AGNs have $\sigma_{\rm [OIII]}$ smaller
by 0.1 dex than that of radio-quiet AGNs with similar $L_{\rm
HOST}$, the narrower $\sigma_{\rm [OIII]}$ for radio-loud AGNs is
the main reason to cause this offset. However, in our whole sample,
we found the median of $\rm \Delta[OIII]$ is about $-300~km~s^{-1}$,
that is about 0.2 dex in narrow line [OIII] width for radio-loud
AGNs comparing to that one in nearby normal galaxies. If the
radio-loud AGNs follow the $M_{\rm BH}-\sigma_{\rm*}$ relation in
nearby normal galaxies, then the real $\sigma_{\rm *}$ in radio-loud
AGNs should be more than 0.2 dex larger than that inferred from the
[OIII] width.


Our sample is simply compiled from literatures, so selection effects
may influence our results.
~If the deviation in $M_{\rm BH}-\sigma_{\rm [OIII]}$ for radio-loud
AGNs was mainly caused by selection effects, which means that the
radio-loud AGNs with low black hole masses and large narrow line
[OIII] width might be lost. However, Laor (2000) studied the black
hole mass in Palomar-Green quasar sample, and pointed out nearly all
PG quasars with $M_{\rm BH}>10^{9}M_{\rm \odot}$ are radio-loud,
while quasars with $M_{\rm BH}<3\times10^{8}M_{\rm \odot}$ are
practically all radio-quiet.
~Therefore there is no this possibility that we have missing so much
this kind of radio-loud AGNs with low black hole mass and large
narrow line [OIII] width during our sample collecting, so we think
the deviation may not mainly be caused by selection effects in our
sample. And we need a complete matching sample including radio-loud
and radio-quiet AGNs to get further study for the influence of
selection effects.


\clearpage
\begin{figure}
   \vspace{2mm}
   \begin{center}
   \hspace{3mm}\epsfig{figure=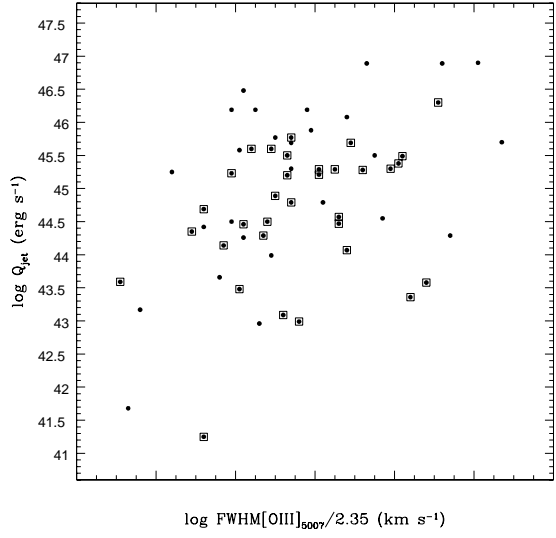,width=80mm,height=80mm,angle=0.0}
   \caption{The relationship between radio jet power and [OIII] line width for radio-loud AGNs.
Solid circles denote all sample and squares added on solid circles
denote black hole mass in range from
   $10^{8.5}M_{\odot}$ to $10^{9.5}M_{\odot}$. }
   \label{Fig:lightcurve-ADAri4}
   \end{center}
\end{figure}

\begin{figure}
   \vspace{2mm}
   \begin{center}
   \hspace{3mm}\epsfig{figure=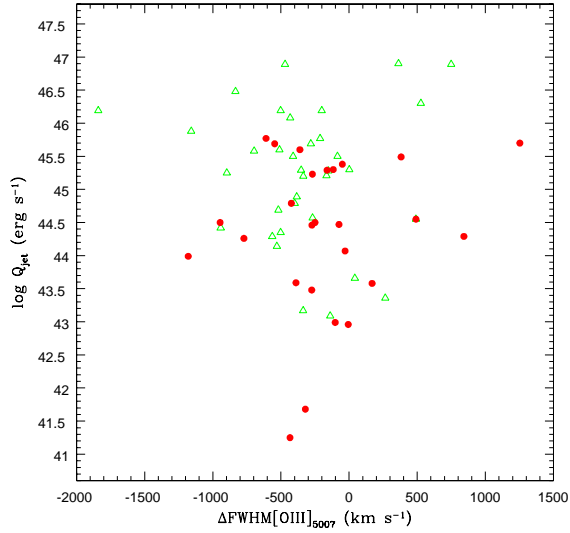,width=80mm,height=80mm,angle=0.0}
   \caption{The relationship between radio jet power and $\rm \Delta[OIII]$ line width
   for radio-loud AGNs. Solid circles denote flat spectrum radio-loud AGNs and open
triangles denote steep spectrum radio-loud AGNs. }
   \label{Fig:lightcurve-ADAri5}
   \end{center}
\end{figure}
\clearpage

The deviation from $M_{\rm BH}-\sigma_{\rm *}$ relation existing in
radio-loud AGNs cannot be explained by beaming effect on optical
continuum luminosity, orientation effect on broad line region and
interaction between radio jet and narrow line region. The redshift
of all objects in our sample are less than 1 because of adopting
[OIII] and $H\beta$ emission line at the same time. It is worth
investigating $M_{\rm BH}-\sigma_{\rm [OIII]}$ relation extends to
higher redshift to avoiding the limitation in sample selection. One
possible explanation is the [OIII] line might not be a good
substitute for stellar velocity dispersion in radio-loud AGNs.

\section{Conclusion}
By estimating the black hole mass eliminating the beaming effects,
we have re-investigated the relationship between black hole mass and
narrow [OIII] line width for a sample of 60 radio-loud and 63
radio-quiet AGNs compiled from literature. Moreover, in order to
know the influence from the radio emission to the narrow line
region, the relationship between radio jet power and the narrow
[OIII] width was also investigated. The main conclusions can be
summarized as follows:
\begin{itemize}
\item{After eliminating beaming effects in optical luminosity and
correcting the orientation effect in broad line width, the $M_{\rm
BH}-\sigma_{\rm [OIII]}$ relation in radio-loud AGNs still deviates
from $M_{\rm BH}-\sigma_{\rm *}$ relation in nearby normal galaxies.
However, it is confirmed again that the radio-quiet AGNs follow the
$M_{\rm BH}-\sigma_{\rm *}$ relation in nearby normal galaxies.}

\item{We find there is no significant correlation between the radio
jet power and the narrow [OIII] line width, which indicating that
the interaction between radio jet and narrow line region is not
obvious. The deviation from $M_{\rm BH}-\sigma_{\rm *}$ in
radio-loud AGNs cannot be explained by interaction in radio jet and
narrow line region.}

\item{The $M_{\rm BH}-\sigma_{\rm [OIII]}$ in radio-loud AGNs deviated
from that relationship in nearby normal galaxies might not be caused
by the uncertainty from black hole mass calculation, indicating that
the narrow [OIII] line width might not be a good substitute for the
stellar velocity dispersion. Another possible explanation is the
radio-loud AGNs might not follow the same $M_{\rm BH}-\sigma_{\rm
*}$ relation as normal galaxies and radio-quiet AGNs, and further
investigation are needed.}

\end{itemize}

\begin{acknowledgements}
We are grateful to Xinwu Cao and Minfeng Gu for helpful discussions.
We thank the anonymous referee for insightful comments and
constructive suggestions. This work is supported by NSFC under grant
10373019 and 10333020. This research has made use of the NASA/ IPAC
Extragalactic Database (NED), which is operated by the Jet
Propulsion Laboratory, California Institute of Technology, under
contract with the National Aeronautics and Space Administration.
\end{acknowledgements}

\clearpage
\begin{table*}
\begin{center}
\begin{small}
\caption{\label{lines}$M_{\rm bh}$ and $\sigma_{[\rm OIII]}$ for our
sample. Col. (1): Object name. Col. (2): Type. Col. (3): Redshift.
Col. (4):log of the black hole mass in units of solar mass. Col.
(5): reference for FWHM of H$\beta$. (6): reference for H$\beta$
luminosity. Col. (7): log of the bulge velocity dispersion derived
from FWHM of [OIII] line in units of $erg~ s^{-1}$. Col. (8):
reference for FWHM of [OIII] line.}
\begin{tabular}{lcccccccc}
\hline\hline
name&type&z&$M_{\rm BH}$ &Refs. &Refs. &$\sigma_{[\rm OIII]}$&Refs.\\
(1)& (2)& (3)&(4)&(5)&(6)&(7)&(8)\\
\hline
0003$+$15 & SS &     0.450  &   9.22  &BG92  & S81 & 2.17 & H84   \\
 0003$+$19 & RQ &     0.026  &   7.51  &S99   & S81 & 2.08 & W92a  \\
 0007$+$10 & FS &     0.089  &   8.93  &B96   & S81 & 2.12 & D88   \\
 0026$+$12 & RQ &     0.142  &   8.05  &BG92  & S81 & 2.02 & D88   \\
 0046$+$31 & RQ &     0.015  &   8.48  &T95   & S81 & 2.19 & W92a  \\
 0049$+$17 & RQ &     0.064  &   8.27  &BG92  & BG92& 2.42 & S89b  \\
 0050$+$12 & RQ &     0.061  &   8.54  &M96   & M96 & 2.81 & M96   \\
 0052$+$25 & RQ &     0.155  &   7.72  &BG92  &SM87 & 2.37 & S89b  \\
 0056$-$00 & FS &     0.717  &   8.55  &B96   &JB91a& 2.62 & B96   \\
 0109$-$38 & RQ &     0.012  &   7.50  &MT98  &MT98 & 2.16 & MT98  \\
 0119$+$22 & RQ &     0.053  &   6.82  &M92   &M92  & 1.91 & F82   \\
 0119$-$01 & RQ &     0.054  &   8.40  &C91   &S81  & 2.41 &D88    \\
 0133$+$20 & SS &     0.425  &   9.83  &JB91b &JB91a& 2.55 &CB96   \\
 0134$+$32 & SS &     0.367  &   8.77  &B96   &JB91a& 2.71 &B96    \\
 0159$-$11 & FS &     0.669  &   9.52  &B96   &O84  & 2.42 &B96    \\
 0205$+$02 & RQ &     0.156  &   8.45  &MD01  &SM87 & 2.34 &GW94   \\
 0211$-$01 & RQ &     0.028  &   7.81  &S99   &S99  & 2.23 &D88    \\
 0232$-$09 & RQ &     0.043  &   8.66  &W92b  &W92b & 2.13 &W92a   \\
 0238$+$06 & RQ &     0.026  &   7.45  &S90   &S90  & 2.24 &W92a   \\
 0316$+$41 & FS &     0.018  &   7.99  &L96   &L96  & 2.26 &M96    \\
 0403$-$13 & FS &     0.571  &   9.10  &M96   &M96  & 2.29 &N95    \\
 0405$-$12 & SS &     0.574  &   9.68  &B96   &M96  & 2.48 &B96    \\
 0414$-$06 & SS &     0.781  &  10.41  &B96   &M96  & 2.39 &B96    \\
 0418$-$55 & RQ &     0.004  &   7.24  &W92b  &W92b & 2.08 &W92a   \\
 0430$+$05 & FS &     0.034  &   8.31  &S99   &S99  & 1.93 &H84    \\
 0434$-$10 & RQ &     0.035  &   7.62  &P82   &S81  & 2.44 &W92a   \\
 0450$-$18 & RQ &     0.058  &   6.18  &S89a  &S89a & 2.43 &S89a   \\
 0513$-$00 & RQ &     0.033  &   8.62  &P82   &S99  & 2.32 &W92a   \\
 0518$+$16 & SS &     0.759  &   8.18  &JB91b &JB91b& 2.72 &GW94   \\
 0521$-$36 & FS &     0.055  &   8.05  &S93   &S93  & 2.74 &S93    \\
 0538$+$49 & SS &     0.545  &   9.70  &L96   &GW94 & 2.81 &L96    \\
 0551$+$46 & RQ &     0.021  &   7.74  &OS82  &S81  & 2.46 &D88    \\
 0609$+$71 & RQ &     0.014  &   8.14  &T95   &S81  & 2.56 &W92a   \\
 0645$+$74 & RQ &     0.020  &   8.10  &S99   &S81  & 2.31 &W92a   \\
 0710$+$11 & SS &     0.768  &  10.87  &B96   &M96  & 2.25 &B96    \\
\hline
\end{tabular}
\end{small}
\end{center}
\end{table*}
\addtocounter{table}{-1}

\begin{table*}
\begin{center}
\begin{small}
\caption{Continued...}
\begin{tabular}{lcccccccc}
\hline\hline
name&type&z&$M_{\rm BH}$ &Refs. &Refs. &$\sigma_{[\rm OIII]}$&Refs.\\
(1)& (2)& (3)&(4)&(5)&(6)&(7)&(8)\\
\hline
 0736$+$01 & FS &     0.191  &   8.67  &B96   &S81  & 2.36 &B96    \\
 0738$+$31 & FS &     0.630  &   9.92  &B96   &S81  & 2.19 &B96    \\
 0738$+$49 & RQ &     0.023  &   8.11  &S99   &S81  & 2.17 &W92a   \\
 0742$+$31 & FS &     0.462  &  10.31  &B96   &M96  & 2.29 &B96    \\
 0754$+$39 & RQ &     0.096  &   8.57  &S89a  &S89a & 2.91 &S89a   \\
 0837$-$12 & SS &     0.200  &   9.29  &B96   &S81  & 2.34 &B96    \\
 0838$+$13 & SS &     0.684  &   8.45  &B96   &M96  & 2.19 &B96    \\
 0903$+$16 & SS &     0.411  &   8.28  &B96   &JB91a& 2.34 &B96    \\
 0921$+$52 & RQ &     0.036  &   7.08  &BG92  &S81  & 2.09 &W92a   \\
 0923$+$39 & FS &     0.699  &   9.55  &B96   &M96  & 2.34 &B96    \\
 0945$+$07 & SS &     0.086  &   7.33  &P84   &O76  & 2.16 &H84    \\
 0953$+$41 & RQ &     0.239  &   9.01  &BG92  &M96  & 2.44 &M96    \\
 0955$+$32 & SS &     0.530  &   7.77  &B96   &M96  & 2.57 &B96    \\
 1004$+$13 & SS &     0.240  &   9.17  &B96   &S81  & 2.30 &B96    \\
 1007$+$41 & SS &     0.611  &   8.89  &B96   &M96  & 2.34 &B96    \\
 1011$-$04 & RQ &     0.058  &   7.21  &BG92  &S96  & 2.23 &WL01   \\
 1020$+$20 & RQ &     0.004  &   7.87  &S99   &S81  & 2.31 &W92a   \\
 1022$+$51 & RQ &     0.045  &   6.97  &BG92  &S81  & 1.53 &W92a   \\
 1028$+$31 & FS &     0.177  &   8.50  &B96   &JB91a& 1.91 &B96    \\
 1048$-$09 & SS &     0.344  &   9.04  &BG92  &SM87 & 2.09 &H84    \\
 1049$-$00 & RQ &     0.357  &   9.45  &BG92  &SM87 & 2.38 &M96    \\
 1100$+$77 & SS &     0.311  &   9.37  &B96   &SM87 & 2.41 &B96    \\
 1103$-$00 & SS &     0.425  &   9.32  &B96   &SM87 & 2.52 &B96    \\
 1116$+$21 & RQ &     0.177  &   8.59  &BG92  &M96  & 2.70 &M96    \\
 1119$+$12 & RQ &     0.049  &   7.42  &BG92  &G99  & 2.30 &F82    \\
 1133$+$21 & RQ &     0.030  &   6.96  &WL01  &G94  & 2.09 &D88    \\
 1137$+$66 & SS &     0.646  &   9.56  &B96   &M96  & 2.38 &B96    \\
 1150$+$49 & FS &     0.334  &   8.73  &B96   &SM87 & 2.22 &B96    \\
 1156$+$29 & FS &     0.729  &   8.65  &B96   &B96  & 2.19 &B96    \\
 1200$+$44 & RQ &     0.002  &   6.34  &L99   &S81  & 2.14 &W92a   \\
 1202$+$28 & RQ &     0.165  &   8.92  &BG92  &M96  & 2.33 &M96    \\
 1216$+$06 & RQ &     0.334  &   9.35  &BG92  &M96  & 2.16 &M96    \\
 1217$+$02 & FS &     0.240  &   8.71  &B96   &JB91a& 2.21 &B96    \\
 1219$+$75 & RQ &     0.070  &   8.29  &C91   &S81  & 2.23 &D88    \\
 1223$+$25 & SS &     0.268  &   9.45  &JB91b &JB91a& 2.27 &H84    \\
 1226$+$02 & FS &     0.158  &   9.46  &BG92  &M96  & 2.61 &M96    \\
 1229$+$20 & RQ &     0.064  &   8.06  &BG92  &S99  & 2.07 &S89b   \\
 1237$-$05 & RQ &     0.009  &   8.35  &P82   &S99  & 2.04 &W92a   \\
 1250$+$56 & SS &     0.320  &   8.52  &B96   &JB91a& 2.33 &B96    \\
\hline
\end{tabular}
\end{small}
\end{center}
\end{table*}
\addtocounter{table}{-1}

\begin{table*}
\begin{center}
\begin{small}
\caption{Continued...}
\begin{tabular}{lcccccccc}
\hline\hline
name&type&z&$M_{\rm BH}$ &Refs. &Refs. &$\sigma_{[\rm OIII]}$&Refs.\\
(1)& (2)& (3)&(4)&(5)&(6)&(7)&(8)\\
\hline
 1253$-$05 & FS &     0.536  &   8.20  &WB86  &M96  & 2.87 &M96    \\
 1302$-$10 & FS &     0.286  &   8.92  &BG92  &M96  & 2.48 &M96    \\
1305$+$06 & SS &     0.599  &   9.35  &B96   &O84  & 2.46 &B96    \\
 1307$+$08 & RQ &     0.155  &   8.30  &BG92  &SM87 & 2.29 &AW91   \\
 1322$-$29 & RQ &     0.014  &   6.73  &D98   &D98  & 1.85 &W92a   \\
 1346$-$30 & RQ &     0.016  &   8.07  &W99   &S81  & 2.38 &S99    \\
 1351$+$64 & RQ &     0.087  &   8.81  &BG92  &M96  & 2.54 &M96    \\
 1351$+$69 & RQ &     0.032  &   8.33  &P82   &S81  & 2.39 &W92a   \\
 1352$+$18 & RQ &     0.158  &   8.60  &BG92  &S89a & 2.78 &S89a   \\
 1353$+$18 & RQ &     0.050  &   7.18  &T95   &T95  & 2.37 &W92a   \\
 1354$+$19 & FS &     0.719  &   9.58  &B96   &B96  & 2.30 &B96    \\
 1355$-$41 & SS &     0.313  &   9.58  &AW91  &T93  & 2.21 &AW91   \\
 1411$+$44 & RQ &     0.089  &   8.11  &BG92  &M96  & 2.42 &M96    \\
 1415$+$25 & RQ &     0.018  &   7.92  &P82   &M96  & 2.30 &M96    \\
 1416$-$12 & RQ &     0.129  &   9.05  &BG92  &S00  & 2.15 &AW91   \\
 1425$+$26 & SS &     0.366  &   9.84  &B96   &S81  & 2.12 &B96    \\
 1434$+$59 & RQ &     0.031  &   8.02  &OS82  &S99  & 2.15 &W92a   \\
 1439$+$53 & RQ &     0.037  &   7.90  &T95   &S81  & 2.20 &W92a   \\
 1440$+$35 & RQ &     0.077  &   7.51  &G99   &S81  & 2.41 &G99    \\
 1444$+$40 & RQ &     0.267  &   8.48  &BG92  &SM87 & 2.36 &M96    \\
 1501$+$10 & RQ &     0.036  &   8.48  &BG92  &G92  & 2.03 &D88    \\
 1510$-$08 & FS &     0.361  &   8.82  &B96   &S81  & 2.28 &B96    \\
 1512$+$37 & SS &     0.371  &   9.70  &B96   &M96  & 2.04 &B96    \\
 1522$+$15 & FS &     0.628  &   8.95  &B96   &B96  & 2.46 &B96    \\
 1534$+$58 & RQ &     0.032  &   8.03  &BG92  &S81  & 2.21 &W92a   \\
 1535$+$54 & RQ &     0.038  &   7.03  &BG92  &S81  & 2.17 &W92a   \\
 1545$+$21 & SS &     0.264  &   9.14  &B96   &S81  & 2.33 &B96    \\
 1612$+$26 & RQ &     0.131  &   8.24  &BG92  &S81  & 2.22 &D88    \\
 1613$+$65 & RQ &     0.129  &   9.30  &BG92  &S90  & 2.42 &D88    \\
 1617$+$17 & RQ &     0.114  &   8.77  &BG92  &BG92 & 1.99 &D88    \\
 1618$+$17 & SS &     0.555  &   9.46  &B96   &M96  & 2.49 &B96    \\
 1622$+$23 & SS &     0.927  &   9.80  &OS82  &S81  & 2.22 &W92a   \\
 1626$+$55 & RQ &     0.133  &   8.48  &BG92  &R85  & 2.54 &S89b   \\
 1641$+$39 & FS &     0.594  &   9.50  &L96   &JB91a& 2.59 &L96    \\
 1704$+$60 & SS &     0.371  &   9.30  &B96   &M96  & 2.24 &B96    \\
 1828$+$48 & SS &     0.691  &   9.86  &L96   &JB91a& 2.53 &L96    \\
 1833$+$32 & SS &     0.058  &   8.64  &C91   &S81  & 2.32 &H84    \\
 1845$+$79 & SS &     0.056  &   8.97  &L96   &S81  & 2.64 &L96    \\
 \hline
\end{tabular}
\end{small}
\end{center}
\end{table*}
\addtocounter{table}{-1}

\begin{table*}
\begin{center}
\begin{small}
\caption{Continued...}
\begin{tabular}{lcccccccc}
\hline\hline
name&type&z&$M_{\rm BH}$ &Refs. &Refs. &$\sigma_{[\rm OIII]}$&Refs.\\
(1)& (2)& (3)&(4)&(5)&(6)&(7)&(8)\\
\hline
1928$+$73 & FS &     0.302  &   9.71  &L96   &M96  & 2.22 &L96    \\
 1939$-$10 & RQ &     0.005  &   7.48  &R93   &W92b & 1.73 &W92a   \\
 2041$-$10 & RQ &     0.034  &   8.36  &S99   &M96  & 2.34 &W92a   \\
 2130$+$09 & RQ &     0.063  &   7.93  &BG92  &S81  & 2.17 &W92a   \\
 2135$-$14 & SS &     0.200  &   9.12  &B96   &M96  & 2.12 &B96    \\
 2141$+$17 & FS &     0.213  &   9.37  &JB91b &M96  & 2.68 &M96    \\
 2201$+$31 & FS &     0.298  &   9.11  &N95   &M96  & 2.45 &M96    \\
 2214$+$13 & RQ &     0.066  &   8.44  &BG92  &S81  & 2.15 &W92a   \\
 2221$-$02 & SS &     0.056  &   8.41  &WB86  &S81  & 1.96 &T93    \\
 2251$+$11 & SS &     0.323  &   8.99  &BG92  &M96  & 2.41 &M96    \\
 2251$-$17 & RQ &     0.068  &   8.51  &C91   &S81  & 2.38 &S89a   \\

\hline
\end{tabular}
\end{small}
\end{center}
\vskip 1mm References: AW91: Appenzeller \& Wagner (1991). B96:
Brotherton (1996). BG92: Boroson \& Green (1992). C91: Corbin
(1991). CB96: Corbin \& Boroson (1996). D88: Dahari et al. (1988).
D98: Delgado et al. (1998). F82: Feldman et al. (1982). G92: Grijp
et al. (1992). G94: Gonzalez et al. (1994). G99: Grupe et al.
(1999). GW94: Gelderman \& Whittle (1994). H84: Heckman et al.
(1984). JB91a: Jackson \& Browne (1991). JB91b: Jackson \& Browne
(1991). L96: Lawrence et al. L99: Leighly (1999). (1996). M92:
Miller et al. (1992). M96: Marziani et al. (1996). MD01: Mclure \&
Dunlop (2001). MT98: Murayama \& Taniguchi (1998). N95: Nerzer et
al. (1995). O76: Osterbrock et al. (1976). O84: Oke et al. (1984).
OS82: Osterbrock \& Shuder (1982). P82: Peterson et al. (1982). P84:
Peterson et al. (1984). R85: Robertis et al. (1985). R93: Rosenblatt
et al. S81: Steiner (1981). (1993). S89a: Stephens (1989). S89b:
Sulentic (1989). S90: Stirpe (1990). S96: Simpson et al. (1996).
S99: Sergeev et al. (1999). S00: Sulentic et al. (2000). SM87:
Stockton \& Mackenty (1987). T93: Tadhunter et al. (1993). T95: Tran
(1995). W92a: Whittle (1992). W92b: Winkler (1992). W99: Wandel et
al. (1999). WB86: Wills \& Browne (1986). WL01: Wang \& Lu (2001).

\end{table*}
\label{lastpage}

\end{document}